\documentclass{hotnets22}

\usepackage{times}  
\usepackage{hyperref}
\usepackage{titlesec}
\usepackage{lastpage}

\hypersetup{pdfstartview=FitH,pdfpagelayout=SinglePage}
\usepackage{xcolor}
\definecolor{pennblue}{cmyk}{1,.65,0,.3}
\definecolor{pennred}{cmyk}{0,1,.65,.34}

\definecolor{bananamania}{rgb}{0.98, 0.91, 0.71} %
\definecolor{babyblue}{rgb}{0.54, 0.81, 0.94} %

\newcommand{\PAR}[1]{\par\smallskip\noindent\textbf{#1}}

\usepackage{alltt}
\usepackage{wrapfig}
\usepackage{cmtt}
\usepackage{paralist}
\usepackage{lastpage}
\usepackage{times}
\usepackage{tikz}
\usepackage{balance}

\usepackage{caption}
\usepackage{subcaption}
\usepackage{xspace}
\usepackage{comment}

\usepackage{enumitem}

\graphicspath{{./figures/}{../figures/}}

\usepackage{url}

\usepackage{booktabs}
\newcommand{\ra}[1]{\renewcommand{\arraystretch}{#1}}

\usepackage{listings}
\usepackage{multicol}

\usepackage{url,fullpage,cite,epsfig,xspace,mathpazo,kpfonts,multirow,dcolumn,colortbl,wasysym,pifont,xspace}
\usepackage[T1]{fontenc}
 
\begin{document}

\title{NotNets: Accelerating Microservices \\by Bypassing the Network}

\author{Submission \#61}

\author{
	\rm Peter Alvaro\\
	UC Santa Cruz
	\and
	\rm Matthew Adiletta\\
	Intel Corporation
	\and
	\rm Adrian Cockroft\\
	OrionX
	\and
	\rm Frank Hady\\
	Intel Corporation	
	\and
	\rm Ramesh Illikkal\\
	Intel Corporation
	\and
	\rm Esteban Ramos\\
	UC Santa Cruz
	\and
	\rm James Tsai\\
	Intel Corporation
	\and
	\rm Robert Soulé\\
	Yale University
}

\maketitle

\begin{abstract}
Remote procedure calls are the workhorse of distributed systems. However, as software engineering trends, such as micro-services and serverless computing, push applications towards ever finer-grained decompositions, the overhead of RPC-based communication is becoming too great to bear. In this paper, we argue that point solutions that attempt to optimize one aspect of RPC logic are unlikely to mitigate these ballooning communication costs. Rather, we need a dramatic reappraisal of how we provide communication. Towards this end, we propose to emulate message-passing RPCs by sharing message payloads and metadata on CXL 3.0-backed far memory. We provide initial evidence of feasibility and analyze the expected benefits.
 \end{abstract}

\section{Introduction:}

A majority of businesses~\footnote{52\%, according to a 2020
survey~\cite{oreilly}} now use microservice-based architectures for
building large-scale applications. Breaking a system into autonomous services
that communicate via a fixed API allows development teams to work
independently in every sense, implementing their services in any way
they want while interacting across services only at the interface
level.  Microservices also provide scaling benefits along several
dimensions: operators can scale individual services independently to
accommodate load and bottlenecks, while managers can scale development
and support teams for individual services independently.

These organizational and operational advantages come at a profound
cost that is becoming too great to bear.  Even looking beyond the more
shocking recent headlines,~\footnote{e.g., Amazon Prime Video's move
to a ``monolithic'' application that saved them 90\% in infrastructure
costs~\cite{prime}} there is increasing evidence that the fundamental costs of
microservices may not justify their flexibility.  In a typical
microservice-based application, a single request flow may trigger
hundreds or even thousands of remote procedure calls (RPCs), which
incur data serialization, kernel crossing, packet processing, queuing
delays, and myriad other resource costs and sources of
latency. Facebook reports~\cite{sriraman20} that only 40\% of the compute
cycles contribute to processing business logic, with the rest being
spent on communication.

To mitigate these ballooning, communication-related overheads we must
focus on RPC, the workhorse of microservices.  Unfortunately, we wish
to reduce overhead without trading any of the generality that made the
architecture attractive, a tradeoff that seems difficult to
navigate. Many point solutions in the literature target perceived
bottlenecks in the RPC stack, including kernel-bypass networking to
reduce data copies and kernel crossings~\cite{kalia14,su17,dragojevic14,li21,biswas18} and hardware
acceleration of (de)
serialization~\cite{pourhabibi20,wolnikowski21,karandikar21}.
Unfortunately, as we show in Section~\ref{sec:motivation}, modern RPC
stacks are highly sensitive to workload characteristics.  Our
experiments, using both client/server benchmarks and simulation of
entire microservice-based applications, reveal that minor variations
in communication pattern can easily shift the bottleneck from
overheads in data transformation, to kernel network stack, to HTTP
header processing, to transport-level security and load balancing.
Time will not be well-spent on point solutions that target and
accelerate a single perceived bottleneck of the communication stack.
The problem is the communication itself.

We advocate something more disruptive.  Memory has been undergoing a
transformation similar to the one experienced by storage a decade ago.
RDMA began to show how memory semantics might transcend the boundaries
of a host; now technologies such as Compute Express Link~\cite{cxl} (CXL) allow
multiple sockets on multiple nodes to access pooled remote memory via
a fast interconnect.  The next step---sharing the data in the memory
``pool'' among those nodes, at low latency and without queuing
effects---is within reach.
This vision may seem perilously close to distributed shared memory (DSM),
an old idea that, each time it comes back around, is dismissed by the systems community as impractical due to the challenges of scaling coherence and
tolerating faults.  We observe that the semantics of RPC, which involve
the round-trip transfer of (immutable) data and control between agents that are
already assumed to fail independently, places very modest constraints on 
the sharing medium, and requires none of the complexity of general DSM.

In this paper, we present \texttt{Notnets}, a network-bypass strategy
that can be retrofitted to existing RPC frameworks.
\texttt{Notnets} will allow a collection of hosts (with a radix of as many as 512-1024 cores) to use a pool
of CXL-attached memory to transparently implement message-passing
semantics in a way that avoids all of the dominant bottlenecks in the
current RPC stack. By exploiting message passing semantics, it 
 does not require us to assume any underlying
coherence mechanism, making it future-safe. Our initial experiments
suggest that network bypass can improve RPC latency by an order of magnitude.

\section{The Overhead of Microservices}
\label{sec:motivation}

\begin{figure}[t]
 \includegraphics[width=3.3in]{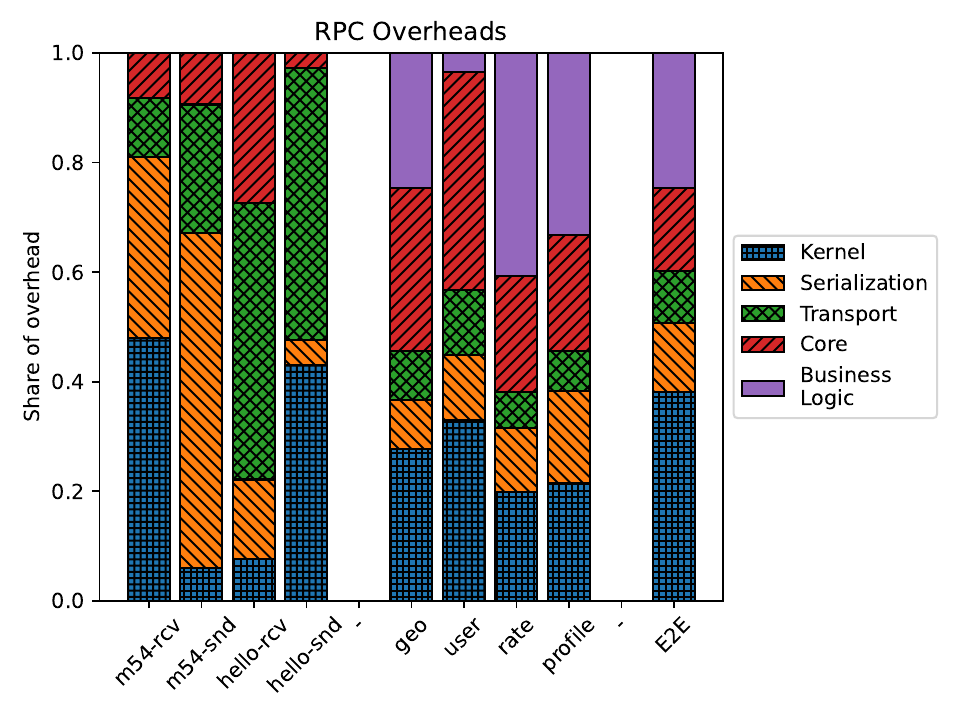} 
\caption{\textbf{Where is time is spent in systems that use RPC?} The first four bars (aHyperprotobenchh, while the second (b) profile individual microservices
  from the \texttt{HotelReservation} application in DeathStarBench. The last bar is the combined profile across all services in the \texttt{HotelReservation}.
}
 \label{fig:microservice-time}
\end{figure}

To better understand what factors contribute to the performance overhead of RPCs, we performed a basic experiment with RPC client-server pairs in which the business logic is to simply ``echo'' the input message.  
The client and server machines were directly connected to a single switch and no other traffic was sent over the network. Thus, network propagation time was negligible. We performed this experiment on a variety of messages, including ``hello world'' from the gRPC tutorial example~\cite{gRPC} 
and \textbf{m54} from Google's Hyperprotobench benchmark\cite{hyperprotobench}, which we report here. We used Intel VTune to capture the call stacks on the gRPC server side, and break down the delay into four categories: 
gRPC Serialization which mainly captures the time spent on serializing/deserializing the messages, gRPC Transport which mainly involves HTTP header processing for the messages, gRPC Core which consists of other gRPC internal processing including setting up a bunch of internal data structures and handling IOs, and kernel stack which is the TCP/IP stack used to receive/send messages.  
The results of this initial experiment are shown in Figure~\ref{fig:microservice-time}a.

These microbenchmarks provide initial evidence that RPC stacks are highly sensitive to workload characteristics such as payload size.  Factors (e.g., kernel networking or data serialization) that dominate in some scenarios
(e.g., bench5, m54 receive and send, respectively) are negligible in others (e.g., helloworld receive and send, respectively).
This suggests that point solutions targeting perceived bottlenecks in the stack are unlikely to bear fruit in mixed workloads.
For example, kernel-bypass networking is an attractive solution to avoid the fixed overheads of kernel crossings and redundant data copies~\cite{biswas18,dragojevic14}, but at the cost of significant complexity. Efforts to port existing RPC frameworks to utilize the flavor of the month in userspace networking will not be well-spent in application regimes in which these fixed overheads are dwarfed by data- and topology-dependent costs in serialization, discovery, and load balancing.  As a second example, the conventional wisdom that serialization costs dominate RPC might lead us to explore point solutions that accelerate serialization with specialized hardware~\cite{pourhabibi20,wolnikowski21,karandikar21}. These efforts might provide only marginal benefit for applications that use large messages with relatively simple serialization logic, and perhaps no benefit at all for applications that favor small messages.

Of course, these microbenchmarks might not reflect the balance of RPC overhead in practice.  What is more, they only study overhead in the RPC stack without considering to what degree this overhead interferes with business logic
in practice.  To get a clearer picture of RPC overhead in the context of a realistic microservice-based application, we performed a second profiling experiment using the social network application in Deathstarbench~\cite{gan19}.
We used the \texttt{HotelReservation} application, consisting of 8 services as well as a persistent backing store, using the \texttt{\scriptsize{mixed-workload\_type\_1}} benchmark.
We ran the workload for 30 minutes, and use golang's \texttt{pprof} package to profile the results,
which are shown in Figure\ref{fig:microservice-time}b.

In addition to plotting time spent in the four RPC categories used in the previous experiment, we include the time spent in user-supplied business logic code.  We show the breakdown for four services (\texttt{geo, user, rate, profile})
as well as the cumulative breakdown across the entire application.  We note first that across all services, only about 25\% of CPU cycles are spent doing useful work.  We are not too surprised to see a lower figure even than Facebook
reported, since the application code in the Deathstarbench applications are very simple.  As we observed in the microbenchmarks, the overhead of different components of the RPC stack is very sensitive to the workload presented by each
individual service.  Finally, RPC overheads are balanced across the entire application (as shown in the \texttt{merged} bar), with time spent in the kernel dominating slightly.

\section{Enabling Trends and Feasibility}

The two observation behind this work are: (i) 
shared, remote memory is possible and almost here, and (ii)
adopting RPC semantics allows 
\texttt{Notnets} to sidestep many of the traditional problems attributed to DSM.

\subsection{Disaggregated and Shared Memory}

Over the past decade, industry has increasingly adopted data center
disaggregation~\cite{klimovic2016,legtchenko17,chung15,
  intelrsd,gao16}, which allows storage and compute resources
to be scaled independently, improving utilization and efficiency.

Now, memory is about to undergo a similar
transition~\cite{mehra2022,lim2009,aguilera18,dragojevic14,gu17}.
This trend is not just an academic exercise. The emerging Compute
Express Link (CXL)~\cite{cxl} open standard and silicon
implementations allow a second, locally-accessible, per-node memory,
enabling DRAM capacity and bandwidth scaling to match CPU scaling.

Early use-cases~\cite{ruan20,zhou22,amaro20} have
focused on scenarios in which remote memory is dynamically allocated
to a particular application, as opposed to shared among a set of
applications.

However, disaggregated shared memory is on the horizon.  The CXL 3.0
standard will enable switch fabric based CXL memory connected to
multiple processors.  Microsoft has shown that 30\% of data capacity
may be be stored in tiered memory resulting in a 10\% overall
reduction in data center DRAM with just 32 notes (sub rack-level)
sharing memory~\cite{li2022pond}.  Optical technology may enable
memory sharing beyond rack-level.  The data center level advantages
achieved by storage in the last decade seem within reach of data
center memory in the coming decade.

This vision may sound like distributed shared memory (DSM) by another
name. DSM, of course, has been studied for
decades~\cite{carriero94,bisiani90,lenoski98,carter91}, and it would
be natural to question why any new attempt to revive this technology
would succeed.

Historically, there have been three main problems that
have hindered the adoption of DSM:

\begin{itemize}[leftmargin=*,noitemsep]
\item \emph{Access Latency.}  Applications that are not written to tolerate non-uniform memory
access latencies have unacceptable performance when some memory accesses are remote.
  
\item \emph{Failures.} Applications are not written to handle partial failures of memory~\cite{schroeder09},
and masking the failure of memory nodes via redundancy~\cite{schneider90,lamport98} incurs
unacceptable costs on the critical path of loads and stores.

\item \emph{Coherence and Synchronization.} Protocols that manage transparent access to copies of shared
data do not scale well~\cite{nagarajan2020,herlihy91}.
\end{itemize}

\subsection{Why RPC is Different}
\texttt{Notnets} can side-step all three of these
problems because it does not need to support arbitrary, transparent
access to shared memory.  All that is required is support for the
basic semantics of RPC, 
which narrowly extends the
standard single-machine procedure call abstraction to provide transfer
of control and data across a computer network. When
the remote procedure is invoked, the caller suspends its execution,
passes the parameters across the network, and executes the procedure
on the callee. When the procedure completes, the results are passed
back to the caller. 
We note these key features of the semantics of RPC:
\begin{enumerate}[leftmargin=*,noitemsep]
\item Access to remote memory is not transparent. Rather,
  it is explicit at the remote method invocation.
\item Agents are already assumed to fail independently.
\item  A chunk of memory is
  only ever explicitly and exclusively owned by the sender or the receiver.
  It is never concurrently shared.
\end{enumerate}

\PAR{Access Latency.}
There are two key arguments to make about performance. First,
distributed applications are already written in such a way that they
expect increased latency for remote access due to the overhead of RPC
communication. Thus, increased latency is not really an issue, when
replacing RPC with shared-memory communication. Second, even if it
were, the latency is likely better with \texttt{Notnets} than with
RPC. The performance gap between inter and intra node communication
has tightened significantly since the early work on DSM. To make the
discussion concrete, we share a few representative performance numbers
for typical systems. The latency for accessing DRAM on a Xeon Skylake
is roughly 62ns~\cite{skylake,xeon}. The latency for accessing a NVMe
SSD (NAND Flash) is about 90$\mu$s~\cite{nvme}.  Ousterhout et
al. report the latency for 1-sided RDMA as 1.4$\mu$s and for a
highly-tuned RPC as 2.4$\mu$s. Thus, we see that although RPC has
100$\times$ higher latency than accessing local DRAM, RPC is also
100$\times$ faster than accessing local SSD. 
Recent work~\cite{gouk2023} reports that DirectCXL memory
pooling achieves around 7$\times$ better performance than RDMA-based
memory pooling.

\PAR{Failures.}
Unlike transparent DSM, the RPC model already assumes that agents can
fail independently and has well-defined semantics in such
contingencies.  Unlike a \texttt{load} or \texttt{store}, a RPC call
can return an error to the caller, due either to an explicit error
return from the server or (in the event that the server node is down
or otherwise unreachable) a client-managed timeout. A typical
microservice is designed to anticipate and mitigate the effects of
failures of services upon which it depends, either by retrying, taking
a fallback path, or supplying a static default.  Hence
\texttt{Notnets} completely sidesteps the problem of transparent
fault-tolerance for remote memory.

\PAR{Coherence and Synchronization.}
The last bugbear of DSM is the performance and scalability of
coherence.  Much like fault tolerance, this concern is off the table
for \texttt{Notnets} because the RPC abstraction does not require
transparent coherence between concurrently-accessed copies of
memory. In RPC, data movement is always explicitly triggered by the
application, at which time the coherence state of the data (the
immutable ``message'' being modeled in shared memory) transitions
atomically from exclusively owned by the sender to read-only and
exclusively owned by the receiver.

\section{Notnets: the Potential}
\label{sec:potential}

Realizing the advantages of our hypothesized shared memory RPC will
not be as simple as waiting for vendors to deliver data center shared
memory.  RPCs provide more than a location abstraction within the data
center, and many mechanisms that cannot simply be bypassed will need
to be rethought to take best advantage of a shared memory
implementation.

Table~\ref{tbl:rpc} identifies the main pieces of functionality for
an RPC call to a microservice-providing host.
In the section below,
we discuss the
low-hanging fruit for which performance advantages can be shown (as we
demonstrate in Section~\ref{sec:eval}) by simply \emph{doing less}, to
stretch goals that will require further design, and potentially, changes the the programming model to
realize.

\begin{table}
  \centering
  \ra{1.2}
  \resizebox{\columnwidth}{!}{%

     \tiny
  \begin{tabular}{@{}ll@{}}
      \toprule

RPC step &  Solution \\
    \midrule

Memory Allocation  &   Arena-based allocator \\
gRPC Data Transform   & Reference-based fast path \\
Load Balancing and DNS Resolution  & Thread allocation \\
Transport Layer Security &   Page-based protection\\
L7 (HTTP) Networking  &  Not needed \\
Kernel Stack  &  Not needed \\
\bottomrule

  \end{tabular}}
  \caption{ The RPC steps for traditional and the hypothesized shared memory RPC.}
\label{tbl:rpc}
\end{table}

\subsection{Low Hanging Fruit}

\PAR{Kernel Stack.}
A traditional RPC incurs two
kernel crossings (single-digit microseconds) and at least as many
copies of the data payload; the use of out-of-process sidecar proxies
(e.g. Envoy), which is common is microservice architectures to handle
cross-cutting concerns including discovery and routing, effectively
doubles both, and network interfaces may incur additional copies. 
\texttt{Notnets} will avoid all kernel crossings.

\PAR{L7 (HTTP) Networking.}
gRPC uses HTTP as a transport mechanism to allow for
streaming requests, i.e., so that applications to avoid the overhead of opening/closing connections.
However, our shared-memory deployment will obviate the need for
traditional network communication, and therefore, make the need for L7
networking unnecessary. We do note, though, that gRPC uses HTTP
headers to pass meta data (e.g., telemetry information) between the
client and server. This mechanism will need to be replaced by a
shared-memory implementation, which should be  straightforward to
implement in a manner similar to passing the RPC payload.

\PAR{gRPC Data Transform, i.e., (De)serialization.}
There is no free lunch, and it will not in general be possible to
simultaneously avoid all serialization costs and support existing
microservice-based applications in their full variety. One of the
touted benefits of using microservices is that they permit decoupling
(and hence autonomy and independent scaling) of development teams
along API boundaries.  This autonomy implies that applications can
(and often will) be polyglot, in the sense that cooperating services
are implemented in separate languages, frameworks, and runtime
environments.

While it may be reasonable in practice to make some
assumptions about memory representation (e.g. endianness), some
serialization cost seems fundamental when sharing values (e.g.,
floating point numbers) between caller and callee.
Endpoints written in different
languages and hosted on different platforms could in principle choose
a common representation (e.g., Apache Arrow) for data intended to be
shared as RPC arguments or returns, obviating the need for
(de)serialization.
Further study of
microservice-based applications are required in order to understand
how pervasive polyglot systems are.  In any case, fast paths can be
explored whenever two adjacent services in the call graph share a
common representation, and this can be determined statically.

\subsection{Open Questions}

\PAR{Transport Layer Security.}
On a single server, security (i.e., privacy) is provided via process
isolation; the virtual address space in one process is completely
separate from the virtual address space in a second process. In other
words, process A cannot access memory in process B. The isolation is
enforced by the memory management unit (MMU). In particular, a process
is not able to directly access physical memory, but rather must use
virtual memory addresses that the MMU translates to physical
address. In this way, the MMU can ensure that process A cannot “name”
addresses in process B.

In a distributed setting, multiple processes, which may reside on
physically separate machines, work together to provide the application
functionality. Each of these processes have isolated virtual memory
spaces. To copy data from one address to another, applications have
typically relied on message passing. Transport Layer Security (TLS)
encrypts the data to ensure that the data is kept private while in
transit.

However, with a shared-memory backend for communication, the security
model changes. The communicating processes already share an address
space. This suggests that, rather than end-to-end encryption, a new
mechanism is needed to ensure isolation of the shared memory. There
must be something equivalent to the MMU that ensures that
non-participating processes cannot “name” the addresses in the shared
memory segment.

\PAR{Load Balancing and DNS Resolution.}
Load balancing is needed to spread the workload evenly across the
additional machines. DNS is often used as a mechanism to discover
available peers. Today, many microservice deployments rely on side-car
proxies to interpose on RPC requests and perform load
balancing. However, recent versions of gRPC include support for
“proxyless” service meshes, in which gRPC can directly process
requests form the control plane using the XDS API.  As with
traditional RPC deployments, our shared-memory approach will adopt a
scale-out approach in which we increase capacity by adding additional
servers to the memory pool. Thus, we will need to maintain some of the
same infrastructure to monitor load and determine which CPU to target
for execution. To balance load, we expect to extend gRPC’s “serverless
proxy” functionality to allow for selecting peers from within the
memory pool based on the load information that we collect. We expect
that there will be some performance gains compared to standard
DNS-based distribution of servers, but the extent of that benefit
would need to be experimentally evaluated.

\PAR{Memory Allocation.}
Dynamic memory allocation has
historically been a source of performance overhead. To offset this cost,
systems programmers have long used a technique known as arena-memory
allocation, in which a large contiguous block of memory is allocated
once at initialization time and managed by a custom memory
manager. Indeed, Google’s gRPC software already advises developers to
use an arena memory allocator for better performance.

In the context of \texttt{Notnets}, avoiding overhead due to
memory allocation will require trading off transparency and making
changes to the programming model. We could, for example, 
 adapt the gRPC arena memory allocator to manage memory from
the shared memory segment, rather than from a standard call to
malloc, avoiding the need to copy data back and forth between
shared and process-local memory. Such a change would be minimally invasive to the user code,
since we can maintain the existing API offered by Google’s arena
memory allocator.

\subsection{Discussion}

By cutting the network out of the distributed system, \texttt{NotNets}
may seem to be proposing to boil the ocean. Nevertheless, this
discussion suggests an incremental path. Our initial prototype
transparently short-circuits the overheads of kernel and layer 7
networking and TLS. It makes no attempt to sidestep serialization
overheads, although the next prototype will exploit a fast path when
client and server share a common memory representation (a property
known when servers are deployed).

\section{Evidence of Feasibility}
\label{sec:feasibility}

We now briefly describe our prototype alongside initial evidence of the promise of \texttt{NotNets}.

\subsection{Network Bypass}
\label{sec:proto}

Ultimately, \texttt{NotNets} will emulate message-passing by sharing message payloads and metadata on CXL 3.0-backed far memory. 
While we wait for this (or a similar) technology to become available we are path-finding.
The initial prototype, designed to answer basic questions about required functionality and best-case performance, runs
on a single host
using System V shared memory.  
The communication channel is realized as a circular buffer;
we interpose on GRPC's request/response API, using the ``custom channel'' extension mechanism, to enqueue and dequeue messages at the client and server, respectively.  

In order to future-proof the prototype against changes in the standard we make no assumptions about how the memory is managed on the device side, and make no assumptions about coherence.
This places constraints on our implementation.  For example, a traditional shared queue implementation that uses semaphores or conditional waits to synchronize threads on empty or full queue
conditions if off the table---what host OS would manage this state?
Instead, we implement a simple pull-based model in which both client and server processes poll shared memory mailboxes to synchronize.

The prototype bypasses all communication-related overhead including the TCP and HTTP stacks just as \texttt{Notnets} will.
It also short-circuits away functionality in GRPC related to load balancing, transport-level security,
discovery, and other features that our ultimate solution must somehow address, as we discussed in Section~\ref{sec:potential}.

\subsection{Evaluation}
\label{sec:eval}

\begin{wraptable}{l}{0.22\textwidth}
\footnotesize    
\ra{1.3}
\begin{tabular} { @{}llll@{} }
\toprule
                                 & \textbf{avg}  & \textbf{p99} & \textbf{p999} \\ \midrule
\textbf{http2} & 209 & 507 & 1430 \\
\textbf{notnets} & 30.9 & 97.5 & 420 \\
\textbf{notnets-} & 8.29 & 23.9 & 187 \\
\bottomrule 
\end{tabular}
\caption{Latency in $\mu$s.}
\label{tab:numbers}
\end{wraptable}

Our initial feasibility experiment focuses on end-to-end RPC latency, supporting the intuition that you can do things a lot faster if you do a lot less.
We use a single Google Cloud Platform \texttt{ec-standard-4} host, configured with 4 vCPUs and 16GB memory, to host the server and client, and reproduce
the ``HelloWorld'' experiment reported in Section~\ref{sec:motivation}.  We measure the latency of a trivial ``echo'' RPC end-to-end from invocation to completion
on the client, in three scenarios. \textbf{http2} uses the standard http-based transport of GRPC.  \textbf{notnets} uses the prototype described in Section~\ref{sec:proto}.
\textbf{notnets-} uses the prototype without serializing the message payload.

Table\ref{tab:numbers} reports latencies in microseconds, showing near an order of magnitude performance gain short-circuiting away the RPC.
Bypassing serialization can offer another 3$\times$ improvement.  We should do this.

\section{Conclusion}

Systems engineering is characterized by tradeoffs, and it is a rare and happy day when we can have our cake and eat it too.  Nevertheless, the emerging systems landscape, driven
by other concerns (in this case, saving money by eliminating stranded memory), has offered us a unique opportunity. We can dip our toes into DSM and enjoy \emph{only} its benefits, postponing its downsides
for future research.

\bibliographystyle{abbrv} 
\begin{small}
\bibliography{main}

\begin{thebibliography}{10}

\bibitem{intelrsd}
{Intel Rack Scale Design (Intel RSD)}.
\newblock
  {\url{https://www.intel.com/content/www/us/en/architecture-and-technology/rack-scale-design-overview.html}}.

\bibitem{prime}
Scaling up the prime video audio/video monitoring service and reducing costs by
  90%, 2023.

\bibitem{aguilera18}
M.~K. Aguilera, N.~Amit, I.~Calciu, X.~Deguillard, J.~Gandhi, S.~Novakovi{\'c},
  A.~Ramanathan, P.~Subrahmanyam, L.~Suresh, K.~Tati, R.~Venkatasubramanian,
  and M.~Wei.
\newblock Remote regions: a simple abstraction for remote memory.
\newblock In {\em 2018 USENIX Annual Technical Conference (USENIX ATC 18)},
  pages 775--787, July 2018.

\bibitem{amaro20}
E.~Amaro, C.~Branner-Augmon, Z.~Luo, A.~Ousterhout, M.~K. Aguilera, A.~Panda,
  S.~Ratnasamy, and S.~Shenker.
\newblock Can far memory improve job throughput?
\newblock In {\em Proceedings of the Fifteenth European Conference on Computer
  Systems}, EuroSys '20, 2020.

\bibitem{bisiani90}
R.~{Bisiani} and M.~{Ravishankar}.
\newblock Plus: a distributed shared-memory system.
\newblock In {\em Proceedings. The 17th Annual International Symposium on
  Computer Architecture}, pages 115--124, May 1990.

\bibitem{biswas18}
R.~Biswas, X.~Lu, and D.~K. Panda.
\newblock Accelerating tensorflow with adaptive rdma-based grpc.
\newblock In {\em 2018 IEEE 25th International Conference on High Performance
  Computing (HiPC)}, pages 2--11, 2018.

\bibitem{carriero94}
N.~Carriero, D.~Gelernter, T.~G. Mattson, and A.~H. Sherman.
\newblock The linda{\textregistered} alternative to message-passing systems.
\newblock {\em Parallel Computing}, 20(4):633--655, 1994.

\bibitem{carter91}
J.~B. Carter, J.~K. Bennett, and W.~Zwaenepoel.
\newblock Implementation and performance of munin.
\newblock In {\em Proceedings of the Thirteenth ACM Symposium on Operating
  Systems Principles}, SOSP '91, pages 152--164, 1991.

\bibitem{cxl}
Compute express link.
\newblock \url{https://www.computeexpresslink.org}, 2023.

\bibitem{dragojevic14}
A.~Dragojevi\'{c}, D.~Narayanan, O.~Hodson, and M.~Castro.
\newblock {FaRM: Fast Remote Memory}.
\newblock In {\em 11th {USENIX} Symposium on Networked Systems Design and
  Implementation (NSDI)}, pages 401--414, Apr. 2014.

\bibitem{gan19}
Y.~Gan, Y.~Zhang, D.~Cheng, A.~Shetty, P.~Rathi, N.~Katarki, A.~Bruno, J.~Hu,
  B.~Ritchken, B.~Jackson, K.~Hu, M.~Pancholi, Y.~He, B.~Clancy, C.~Colen,
  F.~Wen, C.~Leung, S.~Wang, L.~Zaruvinsky, M.~Espinosa, R.~Lin, Z.~Liu,
  J.~Padilla, and C.~Delimitrou.
\newblock An open-source benchmark suite for microservices and their
  hardware-software implications for cloud \& edge systems.
\newblock In {\em Proceedings of the Twenty-Fourth International Conference on
  Architectural Support for Programming Languages and Operating Systems},
  ASPLOS '19, page 3–18, 2019.

\bibitem{gao16}
P.~X. Gao, A.~Narayan, S.~Karandikar, J.~Carreira, S.~Han, R.~Agarwal,
  S.~Ratnasamy, and S.~Shenker.
\newblock Network requirements for resource disaggregation.
\newblock In {\em Proceedings of the 12th USENIX Conference on Operating
  Systems Design and Implementation}, OSDI'16, page 249–264, 2016.

\bibitem{gouk2023}
D.~Gouk, M.~Kwon, H.~Bae, S.~Lee, and M.~Jung.
\newblock Memory pooling with cxl.
\newblock {\em IEEE Micro}, 43(2):48--57, 2023.

\bibitem{gRPC}
grpc.
\newblock \url{https://grpc.io}, 2023.

\bibitem{gu17}
J.~Gu, Y.~Lee, Y.~Zhang, M.~Chowdhury, and K.~G. Shin.
\newblock Efficient memory disaggregation with infiniswap.
\newblock In {\em 14th USENIX Symposium on Networked Systems Design and
  Implementation (NSDI 17)}, pages 649--667, Mar. 2017.

\bibitem{herlihy91}
M.~Herlihy.
\newblock Wait-free synchronization.
\newblock {\em ACM Trans. Program. Lang. Syst.}, 13(1):124--149, Jan. 1991.

\bibitem{hyperprotobench}
Hyperprotobench.
\newblock {\url{https://github.com/google/HyperProtoBench}}, 2022.

\bibitem{nvme}
{{Intel NVMe with 3D XPoint Technology chart}}.
\newblock
  {\url{https://www.tomshardware.com/reviews/intel-micron-3d-xpoint-updates,4286.html#p1}},
  2015.

\bibitem{skylake}
{{Intel Skylake}}.
\newblock {\url{https://www.7-cpu.com/cpu/Skylake.html}}, 2019.

\bibitem{xeon}
{{Intel Xeon Processor E7-8893 v3}}.
\newblock
  {\url{https://ark.intel.com/content/www/us/en/ark/products/84688/intel-xeon-processor-e7-8893-v3-45m-cache-3-20-ghz.html}},
  2019.

\bibitem{kalia14}
A.~Kalia, M.~Kaminsky, and D.~G. Andersen.
\newblock Using rdma efficiently for key-value services.
\newblock In {\em Proceedings of the 2014 ACM Conference on SIGCOMM}, SIGCOMM
  '14, page 295–306, 2014.

\bibitem{karandikar21}
S.~Karandikar, C.~Leary, C.~Kennelly, J.~Zhao, D.~Parimi, B.~Nikolic,
  K.~Asanovic, and P.~Ranganathan.
\newblock A hardware accelerator for protocol buffers.
\newblock In {\em MICRO-54: 54th Annual IEEE/ACM International Symposium on
  Microarchitecture}, MICRO '21, page 462–478, 2021.

\bibitem{klimovic2016}
A.~Klimovic, C.~Kozyrakis, E.~Thereska, B.~John, and S.~Kumar.
\newblock Flash storage disaggregation.
\newblock In {\em Proceedings of the Eleventh European Conference on Computer
  Systems}, EuroSys '16, 2016.

\bibitem{lamport98}
L.~Lamport.
\newblock {The Part-time Parliament}.
\newblock {\em ACM TOCS}, 16(2):133--169, May 1998.

\bibitem{legtchenko17}
S.~Legtchenko, H.~Williams, K.~Razavi, A.~Donnelly, R.~Black, A.~Douglas,
  N.~Cheriere, D.~Fryer, K.~Mast, A.~D. Brown, A.~Klimovic, A.~Slowey, and
  A.~Rowstron.
\newblock Understanding {Rack-Scale} disaggregated storage.
\newblock In {\em 9th USENIX Workshop on Hot Topics in Storage and File Systems
  (HotStorage 17)}, July 2017.

\bibitem{lenoski98}
D.~Lenoski, J.~Laudon, T.~Joe, D.~Nakahira, L.~Stevens, A.~Gupta, and
  J.~Hennessy.
\newblock The dash prototype: Implementation and performance.
\newblock In {\em 25 Years of the International Symposia on Computer
  Architecture (Selected Papers)}, ISCA '98, pages 418--429, 1998.

\bibitem{chung15}
C.~Li, H.~Franke, C.~Parris, and V.~Chang.
\newblock Disaggregated architecture for at scale computing.
\newblock In V.~Chang, M.~Ramachandran, G.~B. Wills, R.~J. Walters, V.~Kantere,
  and C.~Li, editors, {\em ESaaSA 2015 - Proceedings of the 2nd International
  Workshop on Emerging Software as a Service and Analytics, Lisbon, Portugal,
  20-22 May, 2015}, pages 45--52, 2015.

\bibitem{li2022pond}
H.~Li, D.~S. Berger, S.~Novakovic, L.~Hsu, D.~Ernst, P.~Zardoshti, M.~Shah,
  S.~Rajadnya, S.~Lee, I.~Agarwal, M.~D. Hill, M.~Fontoura, and R.~Bianchini.
\newblock Pond: Cxl-based memory pooling systems for cloud platforms, 2022.

\bibitem{li21}
T.~Li, H.~Shi, and X.~Lu.
\newblock Hatrpc: Hint-accelerated thrift rpc over rdma.
\newblock In {\em Proceedings of the International Conference for High
  Performance Computing, Networking, Storage and Analysis}, SC '21, 2021.

\bibitem{lim2009}
K.~Lim, J.~Chang, T.~Mudge, P.~Ranganathan, S.~K. Reinhardt, and T.~F. Wenisch.
\newblock Disaggregated memory for expansion and sharing in blade servers.
\newblock In {\em Proceedings of the 36th Annual International Symposium on
  Computer Architecture}, ISCA '09, page 267–278, 2009.

\bibitem{mehra2022}
P.~Mehra and T.~Coughlin.
\newblock Taming memory with disaggregation.
\newblock {\em Computer}, 55(9):94--98, 2022.

\bibitem{nagarajan2020}
V.~Nagarajan, D.~J. Sorin, M.~D. Hill, D.~A. Wood, and N.~E. Jerger.
\newblock {\em A Primer on Memory Consistency and Cache Coherence}.
\newblock 2nd edition, 2020.

\bibitem{oreilly}
{Microservices Adoption in 2020}.
\newblock \url{https://www.oreilly.com/radar/microservices-adoption-in-2020/},
  2020.

\bibitem{pourhabibi20}
A.~Pourhabibi, S.~Gupta, H.~Kassir, M.~Sutherland, Z.~Tian, M.~P. Drumond,
  B.~Falsafi, and C.~Koch.
\newblock Optimus prime: Accelerating data transformation in servers.
\newblock In {\em Proceedings of the Twenty-Fifth International Conference on
  Architectural Support for Programming Languages and Operating Systems},
  ASPLOS '20, page 1203–1216, 2020.

\bibitem{ruan20}
Z.~Ruan, M.~Schwarzkopf, M.~K. Aguilera, and A.~Belay.
\newblock {AIFM}: {High-Performance}, {Application-Integrated} far memory.
\newblock In {\em 14th USENIX Symposium on Operating Systems Design and
  Implementation (OSDI 20)}, pages 315--332, Nov. 2020.

\bibitem{schneider90}
F.~B. Schneider.
\newblock {Implementing Fault-Tolerant Services Using the State Machine
  Approach: {A} Tutorial}.
\newblock {\em {ACM} Computing Surveys (CSUR)}, 22:299--319, Dec. 1990.

\bibitem{schroeder09}
B.~Schroeder, E.~Pinheiro, and W.-D. Weber.
\newblock {DRAM Errors in the Wild: A Large-scale Field Study}.
\newblock {\em PER}, 37(1):193--204, June 2009.

\bibitem{sriraman20}
A.~Sriraman and A.~Dhanotia.
\newblock Accelerometer: Understanding acceleration opportunities for data
  center overheads at hyperscale.
\newblock ASPLOS '20, page 733–750, 2020.

\bibitem{su17}
M.~Su, M.~Zhang, K.~Chen, Z.~Guo, and Y.~Wu.
\newblock Rfp: When rpc is faster than server-bypass with rdma.
\newblock In {\em Proceedings of the Twelfth European Conference on Computer
  Systems}, EuroSys '17, page 1–15, 2017.

\bibitem{wolnikowski21}
A.~Wolnikowski, S.~Ibanez, J.~Stone, C.~Kim, R.~Manohar, and R.~Soul\'{e}.
\newblock Zerializer: Towards zero-copy serialization.
\newblock In {\em Proceedings of the Workshop on Hot Topics in Operating
  Systems}, HotOS '21, page 206–212, 2021.

\bibitem{zhou22}
Y.~Zhou, H.~M.~G. Wassel, S.~Liu, J.~Gao, J.~Mickens, M.~Yu, C.~Kennelly,
  P.~Turner, D.~E. Culler, H.~M. Levy, and A.~Vahdat.
\newblock Carbink: {Fault-Tolerant} far memory.
\newblock In {\em 16th USENIX Symposium on Operating Systems Design and
  Implementation (OSDI 22)}, pages 55--71, July 2022.

\end{thebibliography}
\end{small}

\end{document}